\documentclass[twocolumn]{aastex63}
\usepackage{bm}
\usepackage{url}
\usepackage{amsmath}
\usepackage{color}
\usepackage{graphicx}

\received{}
\revised{}
\accepted{}
\submitjournal{ApJ}

\shorttitle{Differential Settling in the HL Tau Disk}
\shortauthors{Ueda et al.}
\begin{document}

\title{
Impact of Differential Dust Settling on the SED and Polarization: Application to the Inner Region of the HL Tau Disk
%Differential Settling in the inner region of the HL Tau Disk
%Differential Settling in the HL Tau Disk inferred from Multi-wavelength Continuum and Polarimetric Observations
}

\correspondingauthor{Takahiro Ueda}
\email{takahiro.ueda@nao.ac.jp}

\author[0000-0003-4902-222X]{Takahiro Ueda}
\affil{National Astronomical Observatory of Japan, Osawa 2-21-1, Mitaka, Tokyo 181-8588, Japan}

\author[0000-0003-4562-4119]{Akimasa Kataoka}
\affil{National Astronomical Observatory of Japan, Osawa 2-21-1, Mitaka, Tokyo 181-8588, Japan}

\author[0000-0002-8537-9114]{Shangjia Zhang}
\affil{Department of Physics and Astronomy, University of Nevada, Las Vegas, 4505 S. Maryland Pkwy, Las Vegas, NV 89154, USA}

\author[0000-0003-3616-6822]{Zhaohuan Zhu}
\affil{Department of Physics and Astronomy, University of Nevada, Las Vegas, 4505 S. Maryland Pkwy, Las Vegas, NV 89154, USA}

\author[0000-0003-2862-5363]{Carlos Carrasco-Gonz\'alez}
\affiliation{Instituto de Radioastronom\'{\i}a y Astrof\'{\i}sica (IRyA), Universidad Nacional Aut\'onoma de M\'exico (UNAM)}

\author[0000-0002-5991-8073]{Anibal Sierra}
\affiliation{Departamento de Astronom\'ia, Universidad de Chile, Camino El Observatorio 1515, Las Condes, Santiago, Chile}
%\affiliation{Instituto de Radioastronom\'{\i}a y Astrof\'{\i}sica (IRyA), Universidad Nacional Aut\'onoma de M\'exico (UNAM)}

%% Note that the \and command from previous versions of AASTeX is now
%% depreciated in this version as it is no longer necessary. AASTeX 
%% automatically takes care of all commas and "and"s between authors names.

%% AASTeX 6.2 has the new \collaboration and \nocollaboration commands to
%% provide the collaboration status of a group of authors. These commands 
%% can be used either before or after the list of corresponding authors. The
%% argument for \collaboration is the collaboration identifier. Authors are
%% encouraged to surround collaboration identifiers with ()s. The 
%% \nocollaboration command takes no argument and exists to indicate that
%% the nearby authors are not part of surrounding collaborations.

%% Mark off the abstract in the ``abstract'' environment. 
\begin{abstract}
The polarimetric observations on the protoplanetary disk around HL Tau have shown the scattering-induced polarization at ALMA Band 7, which indicates that the maximum dust size is $\sim 100~{\rm \mu m}$, while the Spectral Energy Distribution (SED) has suggested that the maximum dust size is $\sim$ mm.
To solve the contradiction, we investigate the impact of differential settling of dust grains on the SED and polarization.
If the disk is optically thick, longer observing wavelength traces more interior layer which would be dominated by larger grains.
We find that, the SED of the center part of the HL Tau disk can be explained with mm-sized grains for a broad range of turbulence strength, while $160~{\rm \mu m}$-sized grains can explain barely only if the turbulence strength parameter $\alpha_{\rm t}$ is lower than $10^{-5}$.
We also find that the observed polarization fraction can be potentially explained with the maximum dust size of $1~{\rm mm}$ if $\alpha_{\rm t}\lesssim10^{-5}$, although models with $160~{\rm \mu m}$-sized grains are also acceptable.
However, if the maximum dust size is $\sim3~{\rm mm}$, the simulated polarization fraction is too low to explain the observations even if the turbulence strength is extremely small, indicating the maximum dust size of $\lesssim1$ mm.
The degeneracy between 100 ${\rm \mu m}$-sized and mm-sized grains can be solved by improving the ALMA calibration accuracy or polarimetric observations at (sub-)cm wavelengths. 
\end{abstract}

%% Keywords should appear after the \end{abstract} command.  
%% See the online documentation for the full list of available subject
%% keywords and the rules for their use.

%% From the front matter, we move on to the body of the paper.
%% Sections are demarcated by \section and \subsection, respectively.
%% Observe the use of the LaTeX \label
%% command after the \subsection to give a symbolic KEY to the
%% subsection for cross-referencing in a \ref command.
%% You can use LaTeX's \ref and \label commands to keep track of
%% cross-references to sections, equations, tables, and figures.
%% That way, if you change the order of any elements, LaTeX will
%% automatically renumber them.
%%
%% We recommend that authors also use the natbib \citep
%% and \citet commands to identify citations.  The citations are
%% tied to the reference list via symbolic KEYs. The KEY corresponds
%% to the KEY in the \bibitem in the reference list below. 
\section{Introduction}
Measurement of dust sizes in protoplanetary disks with different ages is a key to understand how and when dust grains grow into larger bodies.
The SED of disks is one of the best ways to constrain the dust size.
If the disk is optically thin at observing wavelengths, the spectral slope of the intensity traces the slope of the dust absorption opacity and hence it allows us to estimate the dust size (e.g., \citealt{Calvet2002,Draine2006}).
In the optically thick regime, the emission is lower than that of the black body because self-scattering reduces the apparent disk brightness (\citealt{Miyake1993,Birnstiel2018,Liu2019,Zhu2019,Carrasco2019,Sierra2020}).
Since the scattering behavior is sensitive to the dust size, we can constrain the dust size from the observed SED even in the optically thick regime \citep{Ueda2020}.

Polarimetric observations at millimeter wavelengths is also useful for constraining dust sizes in protoplanetary disks.
Recent ALMA polarimetric observations have shown that many disks show scattering-induced polarization pattern at the observing wavelength of $\sim$ 1 mm (e.g., \citealt{Bacciotti2018,Hull2018,Dent2019,Sadavoy2019}).
Because self-scattering induces the polarization effectively when the observing wavelength $\lambda$ is comparable to $2\pi a_{\rm max}$ where $a_{\rm max}$ is the maximum dust radius, these observations indicate the prevalence of 100 ${\rm \mu m}$-sized grains in disks even though they should already contain millimeter sized grains.

The disk around the HL Tau young (Class I) star is one of the most intensively studied protoplanetary disks.
The recent multi-wavelength analysis on the HL Tau disk have shown that the inner part of the disk contains $\sim$ mm-sized dust grains \citep{Carrasco2019}.
On the other hand, the ALMA polarimetric observations have shown the transition of the polarization pattern: scattering-induced polarization at ALMA Band 7, alignment-induced polarization at ALMA Band 3 and mixture of them at ALMA Band 6 (\citealt{Stephens2017,Kataoka2017}).
This clear trend indicates that the maximum dust size in the HL Tau disk is $\sim$ 100 ${\rm \mu m}$.

One solution to the contradiction is the differential settling of dust grains caused by disk turbulence (\citealt{Sierra2020,Brunngraber2020,Ohashi2020}, see also \citealt{Liu20} for the polarization of a Class 0 object).
Since the vertical mixing is less efficient for larger grains, larger grains settle more to the mid-plane than smaller grains (\citealt{Dubrulle1995,Youdin2007}).
Therefore, if the disk is optically thick, the shorter observing wavelength traces the more upper layer where smaller grains dominate.
The difference in the observed dust sizes due to the differential settling might make the interpretation of the observations more complicated.

In this paper, we investigate the impact of differential settling on both the SED and polarization of the inner part of the HL Tau disk.
We describe the observational data in Section \ref{sec:2}.
The set up of the numerical simulations are described in Section \ref{sec:3}.
Section \ref{sec:results} gives the comparison between the observations and simulations.
The discussion and summary are in Section \ref{sec:diss} and \ref{sec:summary}.

\section{Observational data} \label{sec:2}
We analyse images of the HL Tau disk observed at several wavelengths obtained in the previous studies. 
In the following analysis, we focus only on the intensity at the center of the observed images, which is the same approach with \citet{Ueda2020}.

For the SED analysis, we use images at the observing wavelength of 0.87 (ALMA Band 7), 1.3 (ALMA Band 6), 2.1 (ALMA Band 4) and 7.9 mm (VLA ${\rm Ka+Q}$), given by \cite{Carrasco2019}.
The synthesized beam size is set to be 0$\farcs$05$\times$0$\farcs$05, corresponding to the spatial resolution of 7.35 au with a distance of HL Tau (147 pc, \citealt{Galli2018}).
The VLA data is corrected by free-free contamination.
For details of the data set, we refer readers to \cite{Carrasco2019}.

\begin{figure}[ht]
\begin{center}
\includegraphics[scale=0.45]{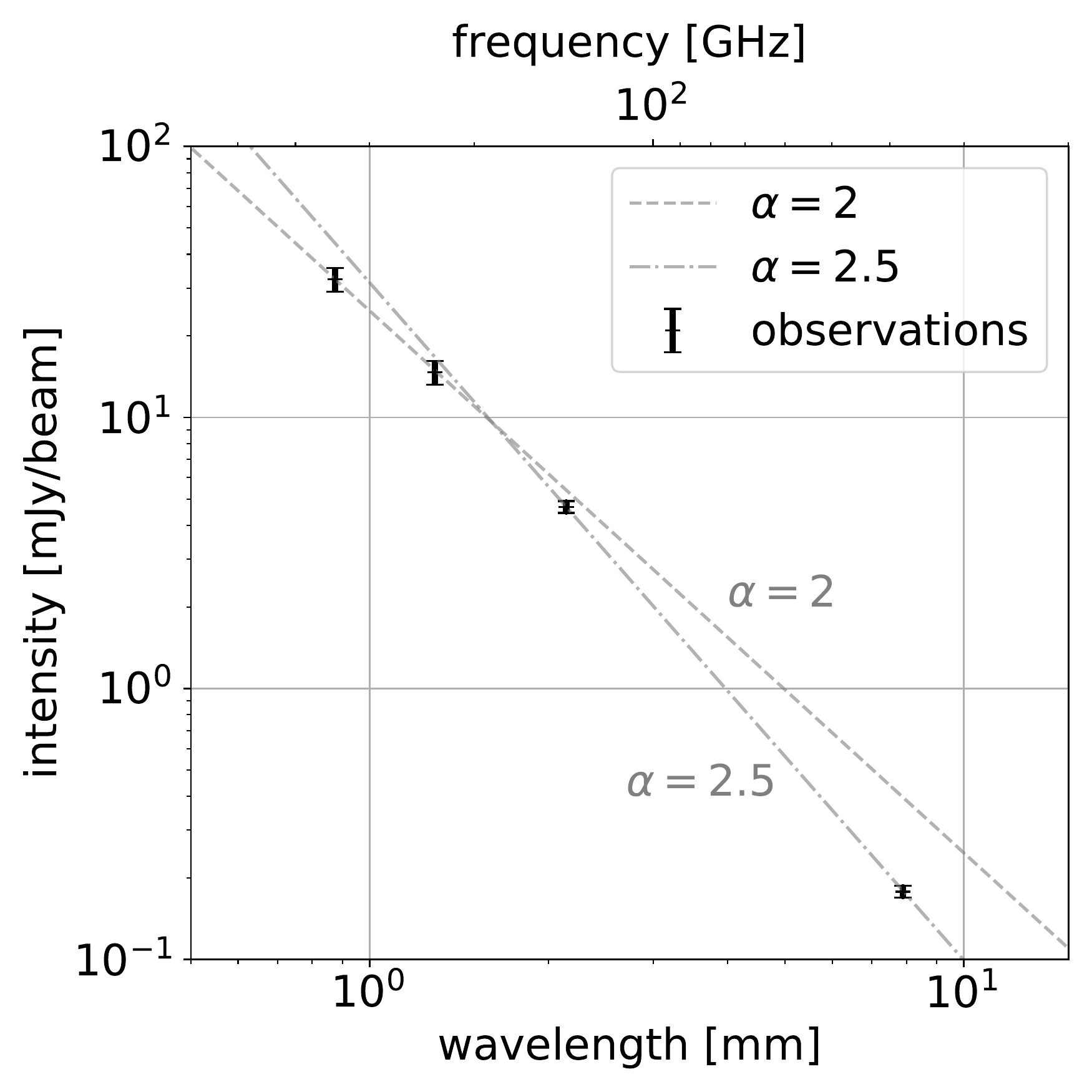}
\caption{
Intensities at the center of the observed images at different observing wavelengths.
The gray dashed and dashed-dotted lines denote the spectral index of 2 and 2.5, respectively.
The uncertainty in the absolute intensity is set to be 10\% for ALMA Band 6 and 7 and 5\% for ALMA Band 4 and VLA observation.
}
\label{fig:sed_obs}
\end{center}
\end{figure}

Figure \ref{fig:sed_obs} shows the intensity at the center of the observed images.
We set the uncertainty in the absolute flux as 10\% for ALMA Band 6 and 7 and 5\% for ALMA Band 4, which are quoted from ALMA official observing guide.
The uncertainty in the absolute flux in the VLA observation is assumed to be 5\%.
These uncertainties might be potentially larger than the nominal values due to e.g., poor weather \citep{Logan2020,Ueda2020}.
At the center of the images, the signal-to-noise is high enough (809, 723, 322 and 60.0 for ALMA Band 7, 6, 4 and VLA observations, respectively) implying that the uncertainty in their intensity is dominated by the flux calibration uncertainty.

The intensity slope between ALMA Band 6 and 7 follows a spectral index of 2.
However, the intensity at ALMA Band 4 is below the intensity extrapolated from the intensity at shorter wavelengths with a spectral slope of 2.
This indicates scattering reduces the intensity at ALMA Band 4 more effectively than at ALMA Band 6 and 7.
The deviation from the spectral slope of 2 is significant at the VLA wavelength and the spectral index between $\lambda=2.1$ mm and 7.9 mm is $\sim$ 2.5.

For the polarization analysis, we use images at the observing wavelength of 0.87 (ALMA Band 7), 1.3 (ALMA Band 6) and 3.1 mm (ALMA Band 3), which are given by \citet{Kataoka2017} and \citet{Stephens2017}.
The angular resolution is
0$\farcs$44$\times$0$\farcs$35,
0$\farcs$37$\times$0$\farcs$24,
0$\farcs$51$\times$0$\farcs$41 for the observation at the observing wavelength of 0.87, 1.3 and 3.1 mm, respectively.
For details of the data set, we refer readers to \citet{Kataoka2017} and \citet{Stephens2017}.

\begin{figure}[ht]
\begin{center}
\includegraphics[scale=0.45]{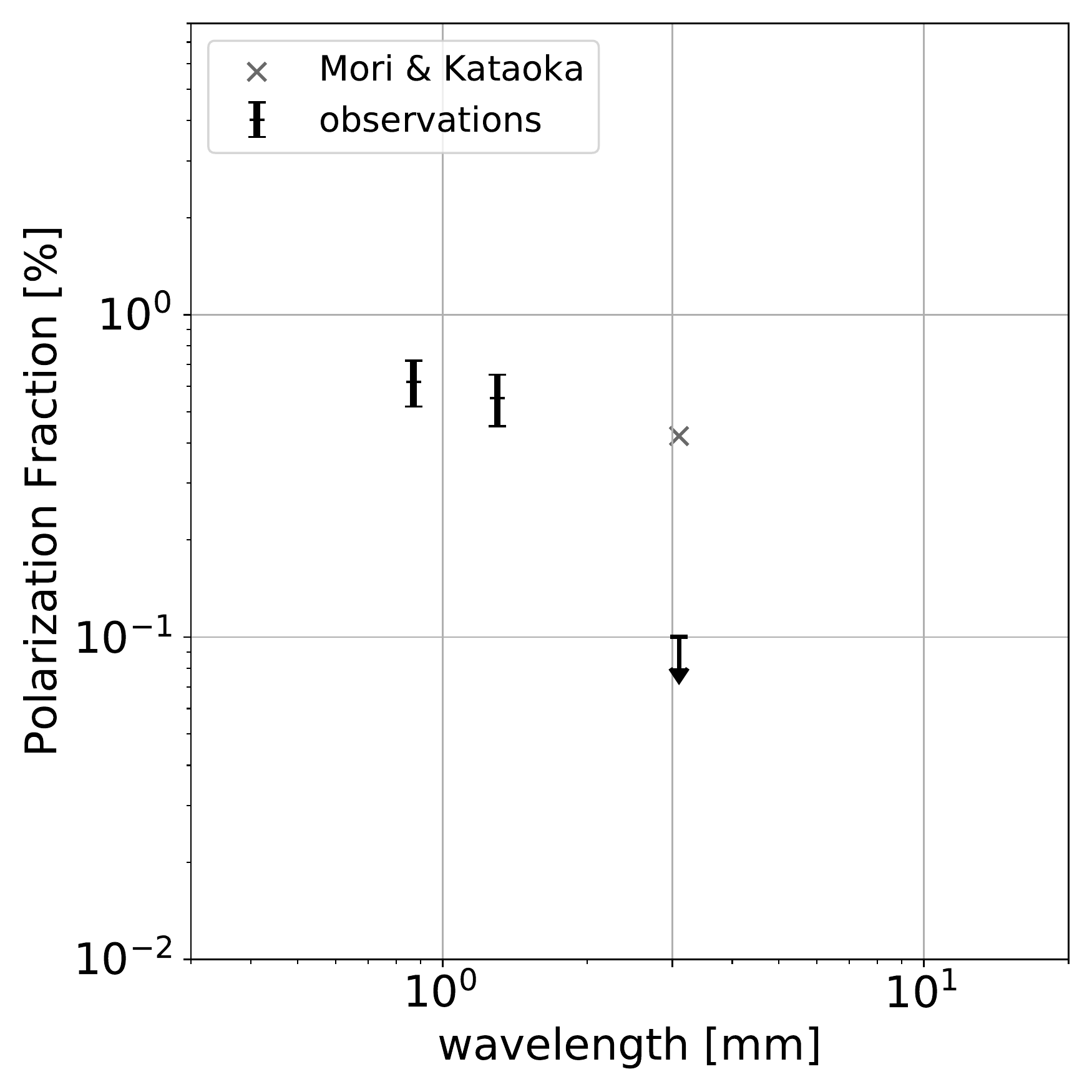}
\caption{
Polarization fraction at the center of the observed images at different observing wavelengths (black crosses).
The uncertainty is set to be 0.1\%.
Only upper limit is obtained for the ALMA Band 3 observation.
The gray cross shows the polarization component induced by self-scattering suggested by \citet{MK2020}.
}
\label{fig:pol_obs}
\end{center}
\end{figure}

Figure \ref{fig:pol_obs} shows the polarization fraction at the center of the observed images.
The observed polarization fraction at $\lambda=0.87$ and 1.3 mm are $\sim$ 0.5\%. while it is less than 0.1\% at $\lambda=3$ mm.
At $\lambda=3$ mm, we have only an upper limit on the polarization fraction since the polarization has not been detected at the center part of the disk.
We set potential errors in the polarization fraction as 0.1\% which corresponds to the ALMA instrumental error.
As shown in \citet{Kataoka2017} and \citet{Stephens2017}, the measured rms in the intensity at the center part of the disk is well below 0.1\%.

The observed polarization might not be originating from only self-scattering.
If dust grains are aligned with the external field such as a magnetic or radiation field, the emission from the aligned grains is polarized and cancel out the scattering-induced polarization (e.g., \citealt{Stephens2017}).
\citet{MK2020} showed that the observed low polarization degree at $\lambda=3$ mm can be explained by the combination of the 0.4\% polarization due to self-scattering and 0.55\% polarization due to dust alignment which has a polarization vector perpendicular to that of self-scattering.
Therefore, for reference, we also plot the self-scattering component suggested by \citet{MK2020} on Figure \ref{fig:pol_obs} with a gray cross.

\section{Radiative transfer simulations} \label{sec:3}
In order to investigate the properties of the inner part of the HL Tau disk, radiative transfer simulations are performed with the Monte Carlo radiative transfer code RADMC-3D \citep{RADMC}.
In this section, we describe how we model the disk in the simulations.

In our simulations, the dust surface density is assumed to follow a power law profile with a gaussian-like gap:
\begin{eqnarray}
\Sigma_{\rm d}= \Sigma_{\rm 0}\left( \frac{r}{\rm 1~au} \right)^{-0.5}\left[ 1-d_{\rm g}\exp{\left\{-\left( \frac{r-r_{\rm g}}{w_{\rm g}} \right)^{2} \right\}}
\right],
\label{eq:surfacedensity}
\end{eqnarray}
where $r$ is the mid-plane distance from the central star and $\Sigma_{\rm 0}$ is the dust surface density at 1 au.
We choose the power-law index of 0.5 for the dust surface density because the intensity profile of the inner region of the disk needs a relatively flat density profile.
We also run some simulations with a power-law index of 1.0 and 1.5 and found that the index of 0.5 is more preferable.
This flat profile is only valid for the inner region where we focus on ($\lesssim20$ au).
Although we focus only on the intensity at the center of the disk and do not focus on the radial profile, we consider the first gap which is located at $r_{\rm g}=12~{\rm au}$ with a depth of $d_{\rm g}$ and width of $w_{\rm g}$.

Using the dust surface density, the dust volume density $\rho_{\rm d}$ is calculated as
\begin{eqnarray}
\rho_{\rm d}=\frac{\Sigma_{\rm d}}{\sqrt{2\pi}h_{\rm d}}\exp{ \left(-\frac{z^{2}}{2h_{\rm d}^{2}}\right) },
\label{eq:volumedensity}
\end{eqnarray}
where $z$ is the vertical height from the mid-plane and $h_{\rm d}$ is the scale height of the dust disk.

\begin{figure}[ht]
\begin{center}
\includegraphics[scale=0.58]{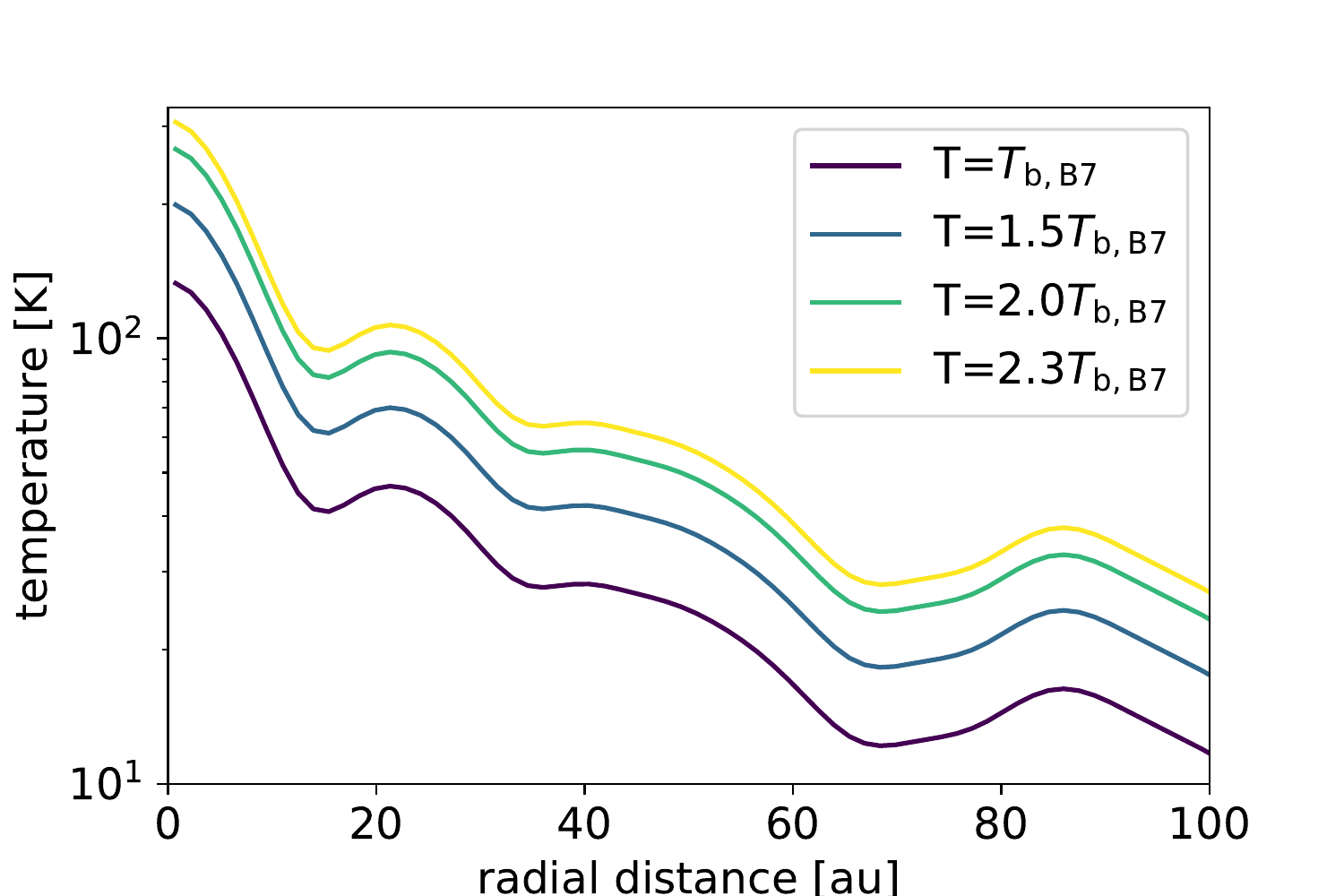}
\caption{Mid-plane temperature profile for different $\epsilon$ values.
}
\label{fig:bt}
\end{center}
\end{figure}

The radial temperature profile is given by the brightness temperature in the Rayleigh-Jeans limit at Band 7, $T_{\rm b,B7}(r)$:
\begin{eqnarray}
T(r)=\epsilon T_{\rm b,B7}(r)=\epsilon\frac{c^{2}}{2k_{\rm b}\nu^{2}_{\rm B7}}I_{\nu_{\rm B7}}(r)
\label{eq:bt}
\end{eqnarray}
where $\nu_{\rm B7}$ and $I_{\nu_{\rm B7}}$ are the observing frequency and intensity at ALMA Band 7.
The true temperature would be different from $T_{\rm b,B7}$ even if the disk is optically thick because scattering reduces the observed intensity.
Therefore we introduce a parameter $\epsilon$ to fit the observed intensity at Band 7.
The temperature profiles for different $\epsilon$ values are shown in Figure \ref{fig:bt}.
Although $\epsilon$ would be also a function of radius since dust population would change with radius, we set $\epsilon$ to be constant with radius for simplicity.
The brightness temperature should be lower than the true disk temperature because of the scattering-induced intensity reduction as well as the optical depth effect.

The dust scale height is assumed to be a mixing-settling equilibrium \citep{Dubrulle1995, Youdin2007}:
\begin{eqnarray}
h_{\rm d}=h_{\rm g} \left( 1+ \frac{\rm St}{\alpha_{\rm t}}\frac{1+2{\rm St}}{1+{\rm St}}\right)^{-1/2}, 
\label{eq:dustheight}
\end{eqnarray}
where $h_{\rm g}$ is the gas scale height given as $h_{\rm g}=c_{\rm s}/\Omega_{\rm K}$ and $\alpha_{\rm t}$ is the stress to pressure ratio \citep{shakura}.
The Stokes number ${\rm St}$ is calculated as
\begin{eqnarray}
{\rm St}=\frac{\pi}{2}\frac{\rho_{\rm int}a}{\Sigma_{\rm g}},
\label{eq:st}
\end{eqnarray}
where $\rho_{\rm int}$ is the material density of dust and $a$ is the dust radius.
The gas surface density $\Sigma_{\rm g}$ is assumed to be $1000~(r/{\rm au})^{-0.5}~{\rm g~cm^{-2}}$, where the power-law index is the same as the dust surface density.

At each radial grid, the vertically integrated dust size distribution is assumed to be a power-law distribution ranging from 0.1 ${\rm \mu m}$ to $a_{\rm max}$ with a power-law index of 3.5.
The dust size distribution ranging from $10~{\rm \mu m}$ to $a_{\rm max}$ is logarithmically divided into 15 size bins per decade and grains smaller than $10~{\rm \mu m}$ is represented with a single size-bin.
For each size-bin, Equation \eqref{eq:dustheight} is applied.
Opacities are calculated with using DSHARP dust optical constants published in \citet{Birnstiel2018} (see also \citealt{Henning1996,Draine2003,Warren2008} for the optical constants of each dust component).
The dust grains are assumed to be spherical compact grains with the material density of 1.675 ${\rm g~cm^{-3}}$.
To treat full anisotropic scattering, the M\"{u}eller matrices are calculated using the Mie theory, specifically Bohren-Huffman program \citep{Bohren1983}.
\begin{figure}[ht]
\begin{center}
\includegraphics[scale=0.56]{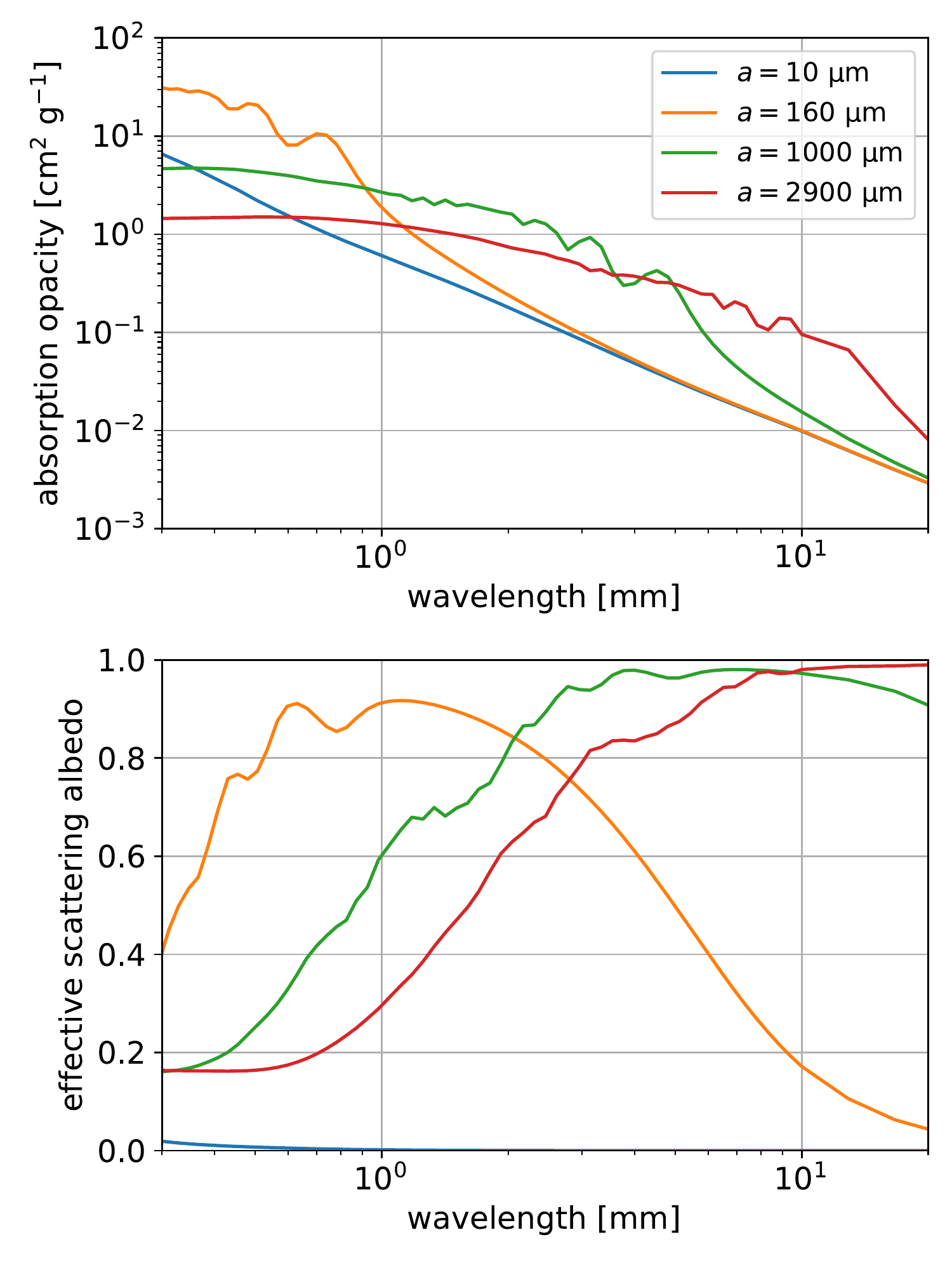}
\caption{Absorption opacity (top) and effective scattering albedo (bottom) for different dust sizes as a function of observing wavelength.
}
\label{fig:opac}
\end{center}
\end{figure}
Figure \ref{fig:opac} shows the absorption opacity $\kappa_{\rm abs}$ and effective scattering albedo $\omega_{\rm eff}$ obtained from our dust model.
The effective scattering albedo is defined as
\begin{eqnarray}
\omega_{\rm eff} = \frac{\kappa_{\rm sca}^{\rm eff}}{\kappa_{\rm abs}+\kappa_{\rm sca}^{\rm eff}},
\label{eq:omegaeff}
\end{eqnarray}
where $\kappa_{\rm sca}^{\rm eff}$ is the effective scattering opacity considering the effect of forward scattering (\citealt{HG1941}; the explicit form is given by e.g., \citealt{Birnstiel2018}).
The sum of the absorption and effective scattering opacity is the extinction opacity $\kappa_{\rm ext}$; $\kappa_{\rm ext}=\kappa_{\rm abs}+\kappa_{\rm sca}^{\rm eff}$.
One of the most important features is that the effective scattering albedo has a peak at $\lambda\sim2\pi a$, which means that the scattering-induced intensity reduction is the most effective at $\lambda\sim2\pi a$ (see \citealt{Ueda2020}).

In the simulation, we use the spherical coordinate and the theta coordinate ranges from $\pi/3$ from $2\pi/3$ where $\pi/2$ corresponds to the mid-plane.
To accurately solve the vertical settling, the calculation domain is divided into 800 grids for the theta direction and  $3\times10^8$ photon packages are used for each simulation.

\section{Results} \label{sec:results}

\subsection{Spectral energy distribution} \label{sec:sed}

In this section, we compare the observed and simulated SED of the center part of the HL Tau disk to see the impact of the differential settling on the SED.

\begin{figure}[ht]
\begin{center}
\includegraphics[scale=0.5]{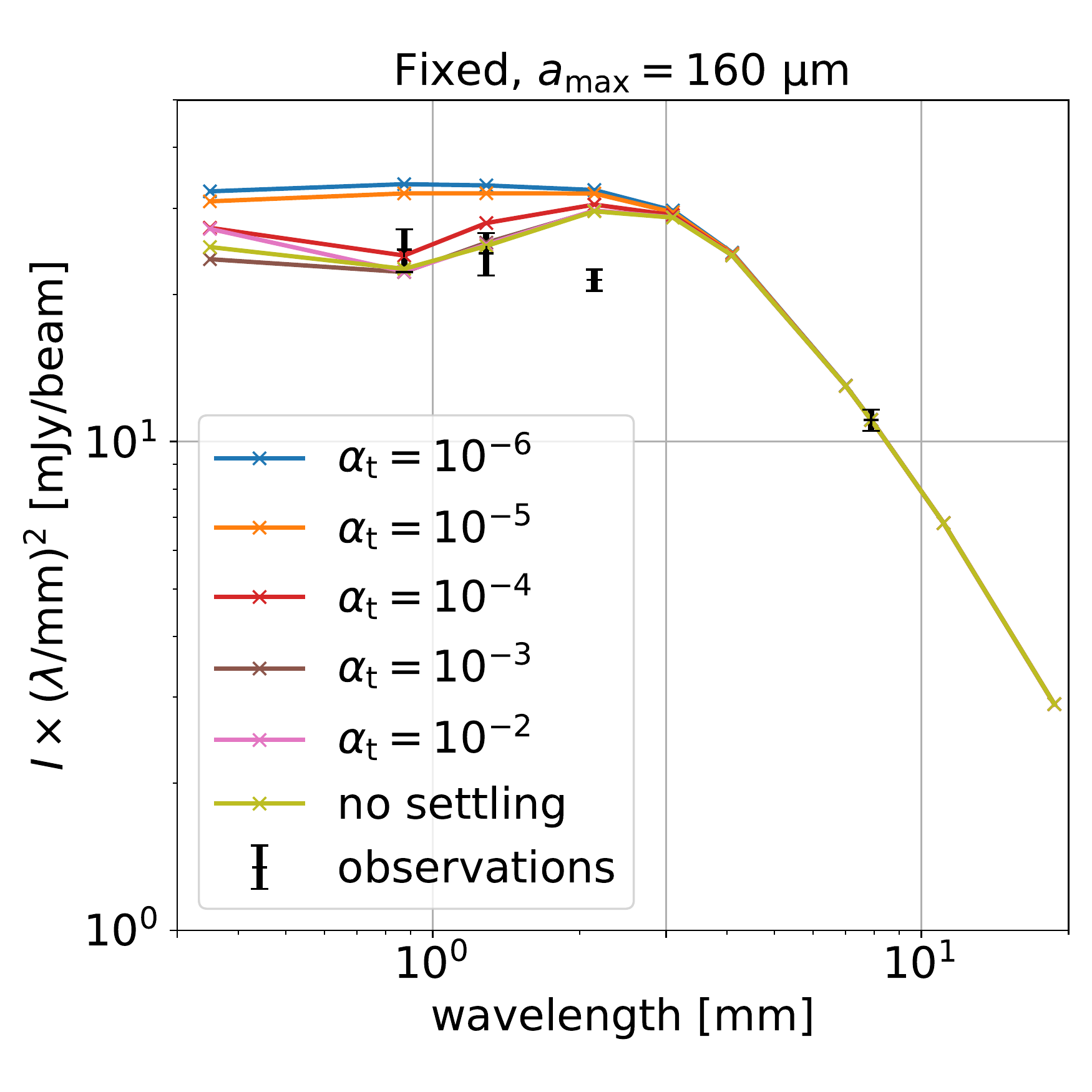}
\caption{
Simulated spectral energy distribution of the center part of the HL Tau disk with the maximum dust size of 160 ${\rm \mu m}$.
$\Sigma_{0}$ and $\epsilon$ are fixed to be $36.7~{\rm g~cm^{-2}}$ and 1.98 in each model, respectively.
}
\label{fig:sed_160}
\end{center}
\end{figure}

\begin{table*}[ht]
  \begin{center}
  \caption{Adopted parameter values}
  \begin{tabular}{ccccccccc}
  \hline
Model & $a_{\rm max}$ & $\alpha_{\rm t}$ & $\Sigma_{0}$ [${\rm g~{cm^{-2}}}$] & $\tau_{\rm 0,8mm}$& $\epsilon$ & $r_{\rm g}$ [au]& $d_{\rm g}$ & $w_{\rm g}$ [au]\\
\hline\hline

& $160~{\rm \mu m}$ & all & 36.7 & 0.59 & 1.98 & 12 & 0.9 & 4.5 \\%\cline{2-9}
Fixed & $1000~{\rm \mu m}$ & all & 24.3 & 8.9 & 2.28 & 12 & 0.97 & 5\\%\cline{2-9}
& $2900~{\rm \mu m}$ & all & 30.0 & 86 & 2.05 & 12 & 0.98 & 9 \\\hline

& & no settling & 36.7 & 0.59 & 1.98 & 12 & 0.9 & 4.5\\%\cline{3-9}
& & $10^{-2}$ & 36.7 & 0.59 & 1.98 & 12 & 0.9 & 4.5\\%\cline{3-9}
& $160~{\rm \mu m}$ & $10^{-3}$ & 38.3 & 0.62 & 1.92 & 12 & 0.9 & 4.5\\%\cline{3-9}
& & $10^{-4}$ & 46 & 0.74 & 1.71 & 12 & 0.9 & 4.5\\%\cline{3-9}
& & $10^{-5}$ & 58.3 & 0.94 & 1.35 & 12 & 0.9 & 4.5\\%\cline{3-9}
& & $10^{-6}$ & 60.0 & 0.96 & 1.32 & 12 & 0.9 & 4.5\\\cline{2-9}

& & no settling & 24.3 & 8.9 & 2.28 & 12 & 0.97 & 5\\%\cline{3-9}
& & $10^{-2}$ & 24.3 & 8.9 & 2.26 & 12 & 0.97 & 5\\%\cline{3-9}
Tuned & $1000~{\rm \mu m}$ & $10^{-3}$ & 23.7 & 8.7 & 2.30 & 12 & 0.97 & 5\\%\cline{3-9}
& & $10^{-4}$ & 26 & 9.5 & 2.15 & 12 & 0.97 & 5\\%\cline{3-9}
& & $10^{-5}$ & 32.7 & 12.0 & 1.75 & 12 & 0.97 & 5\\%\cline{3-9}
& & $10^{-6}$ & 36 & 13.2 & 1.6 & 12 & 0.97 & 5\\\cline{2-9}

& & no settling & 30.0 & 86 & 2.05 & 12 & 0.98 & 9\\%\cline{3-9}
& & $10^{-2}$ & 29.3 & 84.0 & 2.09 & 12 & 0.98 & 9\\%\cline{3-9}
& $2900~{\rm \mu m}$ & $10^{-3}$ & 28.7 & 82.3 & 2.12 & 12 & 0.98 & 9\\%\cline{3-9}
& & $10^{-4}$ & 19.3 & 55.3 & 2.23 & 12 & 0.98 & 9\\%\cline{3-9}
& & $10^{-5}$ & 19.3 & 55.3 & 2.1 & 12 & 0.98 & 9\\%\cline{3-9}
& & $10^{-6}$ & 19.3 & 55.3 & 2.02 & 12 & 0.98 & 9\\\hline
  \end{tabular}
  \end{center}
  \label{table:1}
\end{table*}

Figure \ref{fig:sed_160} shows the simulated SED at the center of the disks with the maximum dust size of 160 ${\rm \mu  m}$ for different $\alpha_{\rm t}$ values. 
The simulated images are convolved with the same beam size as the observations (0$\farcs$05$\times$0$\farcs$05).
For the visual purpose, we multiply the intensity by a square of the wavelength so that the profile is horizontal if the intensity follows the spectral slope of 2 (i.e. optically thick and no-scattering limit).
In Figure \ref{fig:sed_160}, we show the results of Fixed model where the temperature and surface density of the disks are fixed between simulations with different $\alpha_{\rm t}$ values to see the impact of differential settling.
In the Fixed model, the disk parameters are determined so that the simulated SED matches with the observations at $\lambda=0.87$ and 7.9 mm in the no-settling limit.
All adopted parameters in our calculations are summarized in Table \ref{table:1}.

The intensity steeply decreases with $\lambda$ at $\lambda>3~{\rm mm}$ because the disk becomes optically thin. 
In this regime, the intensity does not depend on the turbulence strength since all grains can be seen and hence the vertical distribution does not matter.
In contrast, in the optically thick regime ($\lambda<3~{\rm mm}$), the model with weaker turbulence yields higher intensity.
This is because weaker turbulence makes grains more settle down to the mid-plane, which makes the effective grain size at photosphere smaller.
Since the scattering albedo of $160~{\rm \mu m}$ grains has a peak at $\lambda\sim 1~{\rm mm}$ (Figure \ref{fig:opac}), the SED has a dip at $\lambda\sim 1~{\rm mm}$ due to scattering-induced intensity reduction in the no-settling model.
The depth of the dip at $\lambda\sim 1~{\rm mm}$ is smaller for models with weaker turbulence, because smaller grains has smaller scattering albedo at $\lambda\sim 1~{\rm mm}$.

\begin{figure}[ht]
\begin{center}
\includegraphics[scale=0.45]{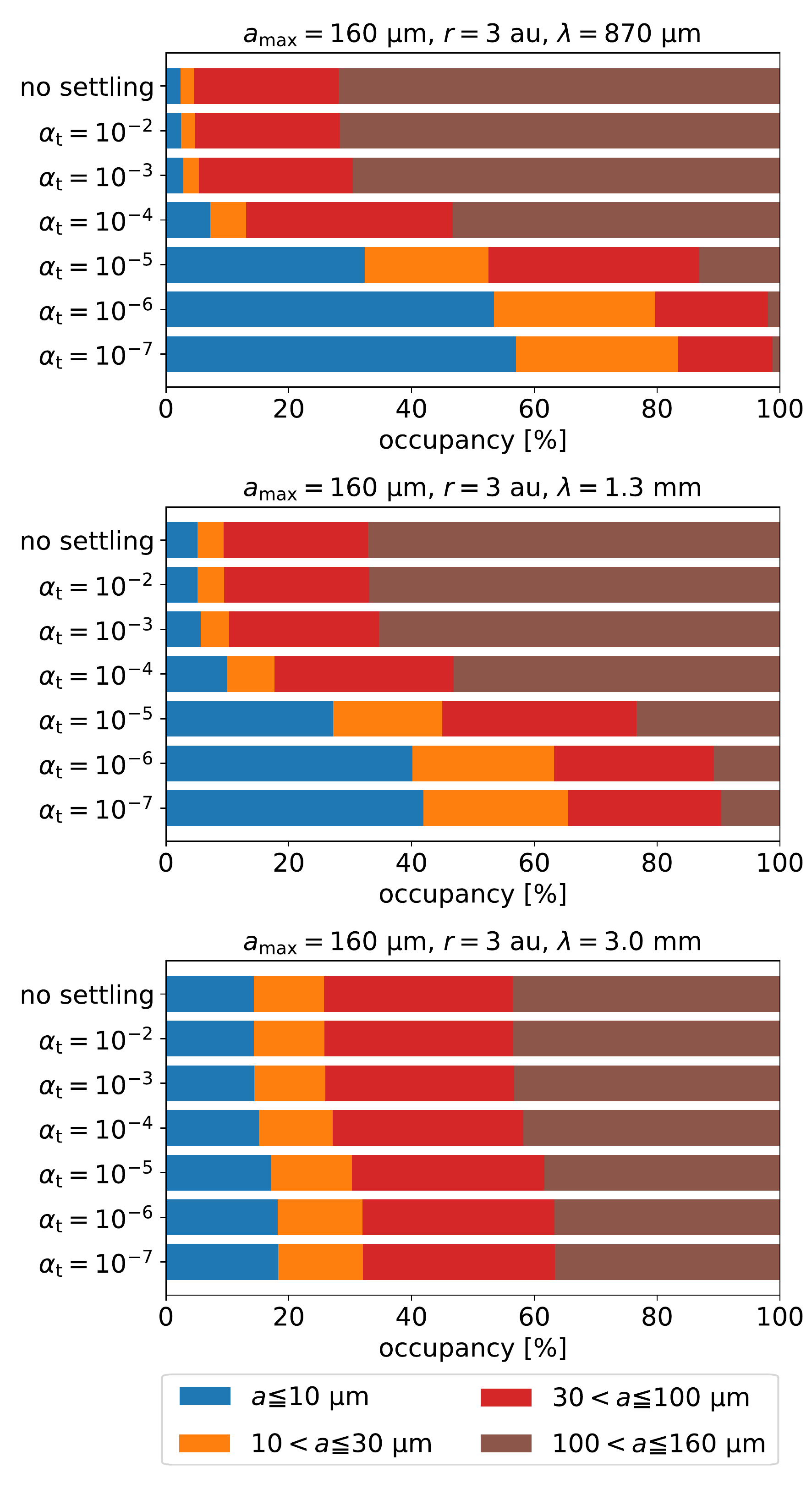}
\caption{Breakdown of vertical extinction optical depths at $\lambda=870~{\rm \mu m}$ (top), 1.3 mm (middle) and 3.0 mm (bottom) from each dust size of Fixed model with $a_{\rm max}=160~{\rm \mu m}$. The optical depths are calculated only with grains above the $\tau=1$ surface. The radial location is 3 au.
}
\label{fig:occupancy-160-tuned}
\end{center}
\end{figure}
To more quantitatively see the impact of differential settling, we plot a breakdown of the extinction optical depth coming from each dust size at radial position of 3 au in Figure \ref{fig:occupancy-160-tuned}.
Since the spatial resolution of the observing beam is $7.35$ au, the intensity at the center of the images mainly traces the radial location at $\sim$ 3 au.
The optical depth is calculated with only grains above the $\tau=1$ surface.
In Figure \ref{fig:occupancy-160-tuned}, we also plot the model with extremely weak turbulence ($\alpha_{\rm t}=10^{-7}$) for reference.
At ALMA Band 7 ($\lambda=870~{\rm \mu m}$), more than 70\% of the observed emission comes from grains with radius of $100<a\leq160~{\rm \mu m}$ in no-settling model, while it is less than 5\% when $\alpha_{\rm t}=10^{-6}$.
In the model with $\alpha_{\rm t}=10^{-6}$, almost the half of the observed emission comes from grains smaller than 10 ${\rm \mu m}$.
In contrast, at ALMA Band 3 ($\lambda = 3~{\rm mm}$), the breakdown is almost the same in all models because the disk is almost optically thin.

\begin{figure*}[ht]
\begin{center}
\includegraphics[scale=0.37]{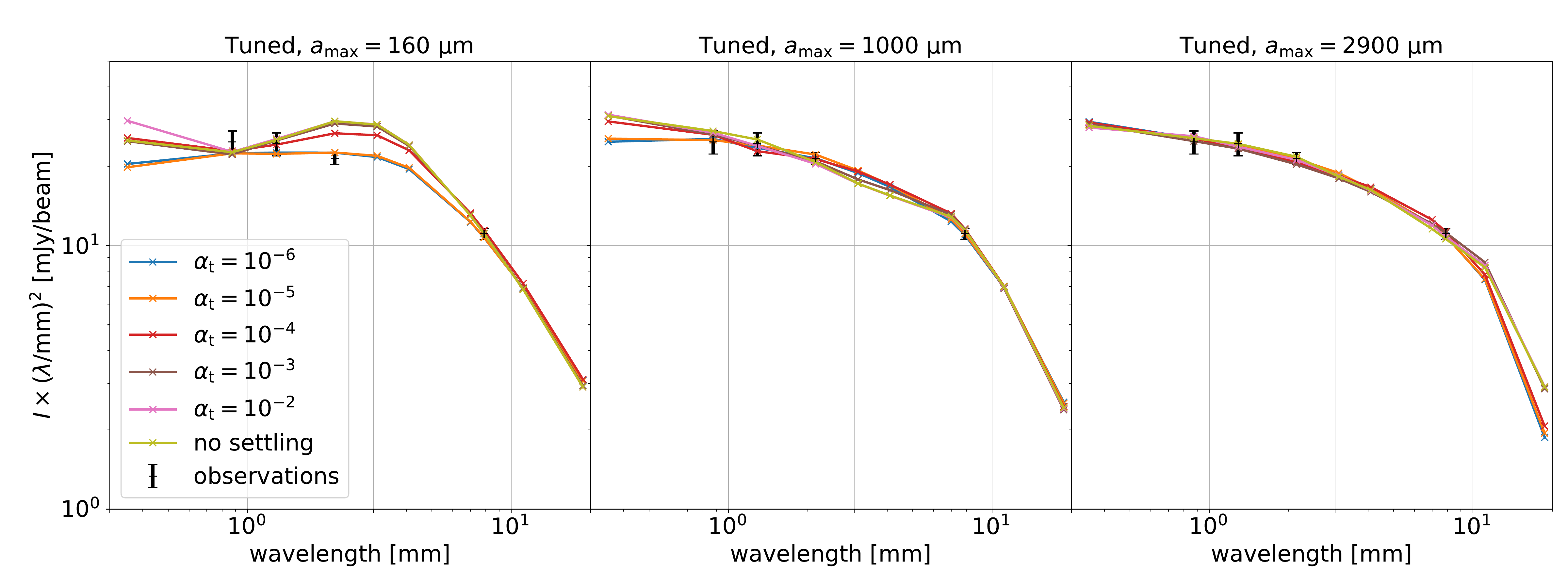}
\caption{
Spectral energy distribution at the center of the HL Tau disk with the maximum dust size of 160 ${\rm \mu m}$ (left), 1000 ${\rm \mu m}$ (center) and 2900 ${\rm \mu m}$ (right).
The temperature and surface density profile are tuned to fit the observed intensity at $\lambda=0.87$ and 7.9 mm in each model.
}
\label{fig:sed_tuned}
\end{center}
\end{figure*}

Figure \ref{fig:sed_tuned} shows the simulated SED for the different maximum dust sizes. 
In Figure \ref{fig:sed_tuned}, we plot the results of Tuned models where the temperature and surface density profile (i.e., $\Sigma_{0}$ and $\epsilon$) are tuned to fit the observed intensity at $\lambda=0.87$ and 8 mm in each simulation.
Basically, the temperature is constrained from the intensity at the shorter wavelength (ALMA Band 7) since the disk would be optically thick.
For a given temperature, the dust surface density is constrained from the intensity at the longer wavelength (VLA ${\rm Ka+Q}$).

Since the different maximum dust sizes have different absorption/scattering opacities (Figure \ref{fig:opac}), the combination of the temperature and optical depth for the observed SED to be reproduced is also different for different dust models. 
For each dust model, we fixed the gap structure for simplicity.
Even though the obtained intensity profile is slightly different from the observed one at the gap region for some $\alpha_{\rm t}$ models, we confirmed that the slight difference in gap structure does not affect the SED and polarization significantly.

We clearly see that the models with the maximum grain size of 160 ${\rm \mu m}$ cannot reproduce the observed intensity profile if $\alpha_{\rm t}>10^{-5}$.
This is because 160 ${\rm \mu m}$-sized grains reduce the intensity efficiently at ALMA Band 7 and hence cannot reproduce the observed low intensity at ALMA Band 4.
If $\alpha_{\rm t}\lesssim10^{-5}$, the effective dust size is very small owing to the settling of large grains and hence the observed intensity profile can be reproduced barely.
The models with $\alpha_{\rm t}\lesssim10^{-5}$ yield the intensity corresponding to the lower limit and the upper limit of the intensity at ALMA Band 7 and 4, respectively.

In contrast, if the maximum dust size is 1 mm or 2.9 mm (Figure \ref{fig:sed_tuned} center and right), the observed intensity profile can be reproduced for a broad range of the turbulence strength because mm-sized grains has high scattering albedo at ALMA Band 4.
This is consistent with the result of \citet{Carrasco2019}.
Although the SED depends on the turbulence strength even in models with mm-sized grains, it is difficult to constrain the turbulence strength from the SED with the given intensity accuracy.

\subsection{Polarization fraction}
As previous studies have shown, polarization should be the most effective when $\lambda\sim2\pi a_{\rm max}$ in the no-settling limit.
However, if differential settling takes place, the polarization behavior would depend on how much large grains exist above the surface where the vertical optical depth is unity.  
In this section, we investigate whether the observed polarization can be explained with mm-sized grains or needs 100 ${\rm \mu m}$-sized grains as previously expected.

\begin{figure*}[ht]
\begin{center}
\includegraphics[scale=0.35]{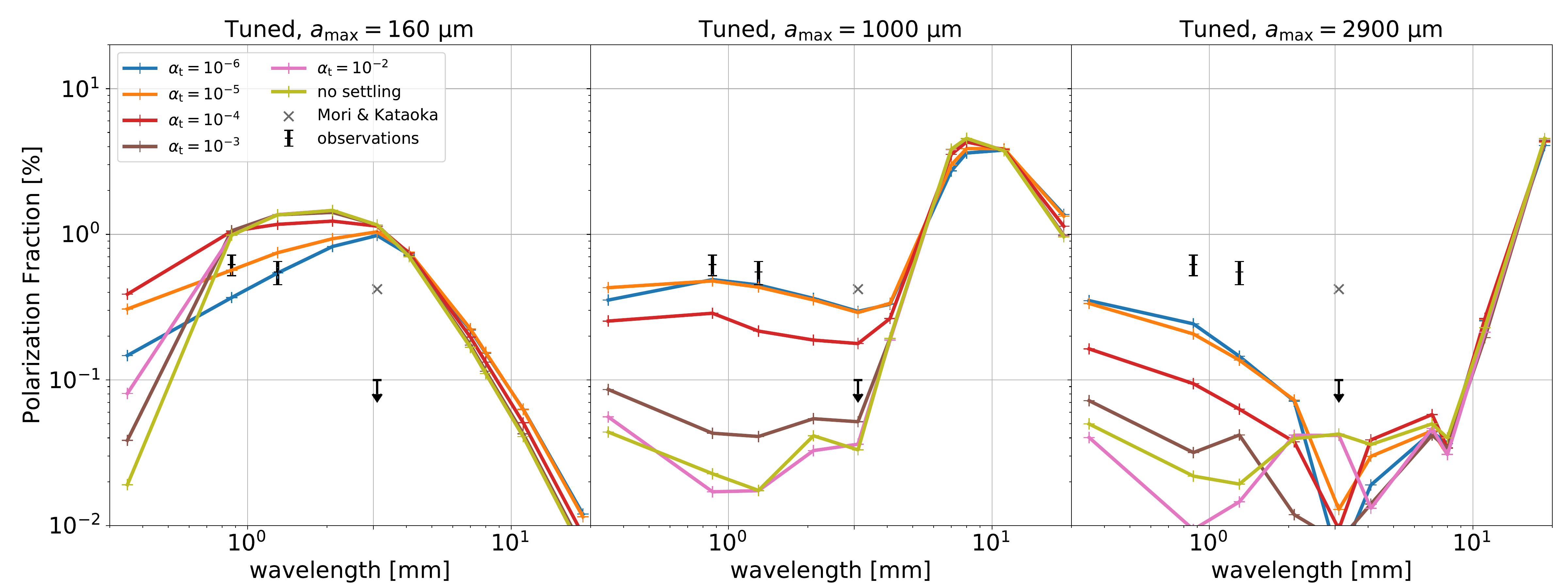}
\caption{
Polarization fraction at the center of the HL Tau disk with the maximum dust size of 160 ${\rm \mu m}$ (left), 1000 ${\rm \mu m}$ (center) and 2900 ${\rm \mu m}$ (right).
The temperature and surface density profile are tuned to fit the observed intensity at $\lambda=0.87$ and 7.9 mm in each model.
The gray cross denotes the polarization fraction induced by self-scattering suggested by \citet{MK2020}.
}
\label{fig:pol_tuned}
\end{center}
\end{figure*}

Figure \ref{fig:pol_tuned} shows the polarization fraction caused by self-scattering at the center of the HL Tau disk. 
The simulated images are convolved with the synthesized beam size of 0$\farcs$3.
It is worth to be noted that the beam size for the polarization analysis is 6 times larger than that for the SED analysis, meaning that the beam in the polarimetric observation averages broader region.

If the maximum dust size is 160 ${\rm \mu m}$, the simulated polarization fraction is $\sim$ 1\% at $\lambda=$1--3 mm for $\alpha_{\rm t}\geq10^{-4}$, which is higher than the observed values.
If $\alpha_{\rm t}=10^{-5}$, the simulated polarization fraction at ALMA Band 7 is consistent with the observation, but higher at ALMA Band 6.
If $\alpha_{\rm t}=10^{-6}$, the simulated polarization fraction at ALMA Band 7 is too small to explain the observation.
The models with $a_{\rm max}=1000~{\rm \mu m }$ yield the polarization fraction consistent with the observations at ALMA Band 6 and 7 when $\alpha_{\rm t}\leq10^{-5}$, while predict very low polarization fraction when $\alpha_{\rm t}\geq10^{-4}$
In contrast to these models, the models with $a_{\rm max}=2900~{\rm \mu m }$ predict very low polarization fraction compared to the observations even if $\alpha_{\rm t}\leq10^{-5}$.
Since the other polarization mechanisms might take place and cancel out the scattering-induced polarization, polarization fraction higher than the observations would be acceptable.
Therefore, we cannot exclude the possibility that the maximum dust size is 160 ${\rm \mu m}$ and the turbulence strength is higher than $10^{-5}$ from the polarization analysis.
All models that potentially account for the observed polarization fraction at ALMA Band 6 and 7 predict polarization fraction higher than the observation at ALMA Band 3, indicating that the other polarization mechanisms are necessary to explain the observed polarization fraction at ALMA Band 3 \citep{MK2020}.

\begin{figure*}[ht]
\begin{center}
\includegraphics[scale=0.31]{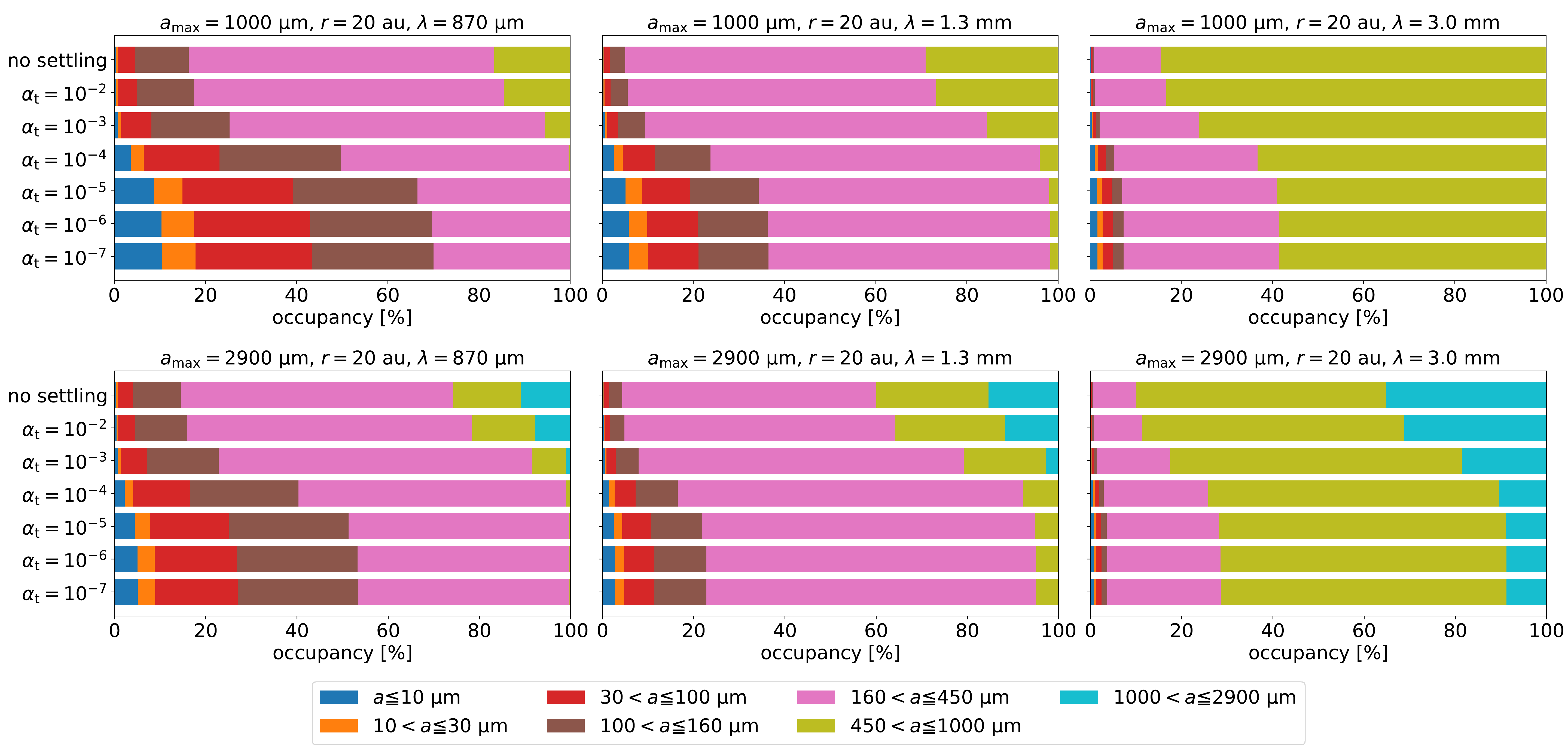}
\caption{Proportion of vertical optical depths at $\lambda=870~{\rm \mu m}$ (top), 1.3 mm (middle) and 3.0 mm (bottom) from each dust size of Fixed model with $a_{\rm max}=1000~{\rm \mu m}$ (top) and $a_{\rm max}=2900~{\rm \mu m}$ (bottom). The optical depths are calculated only with grains above the $\tau=1$ surface. The radial position is 20 au.
}
\label{fig:occupancy-all}
\end{center}
\end{figure*}

\begin{figure*}[ht]
\begin{center}
\includegraphics[scale=0.6]{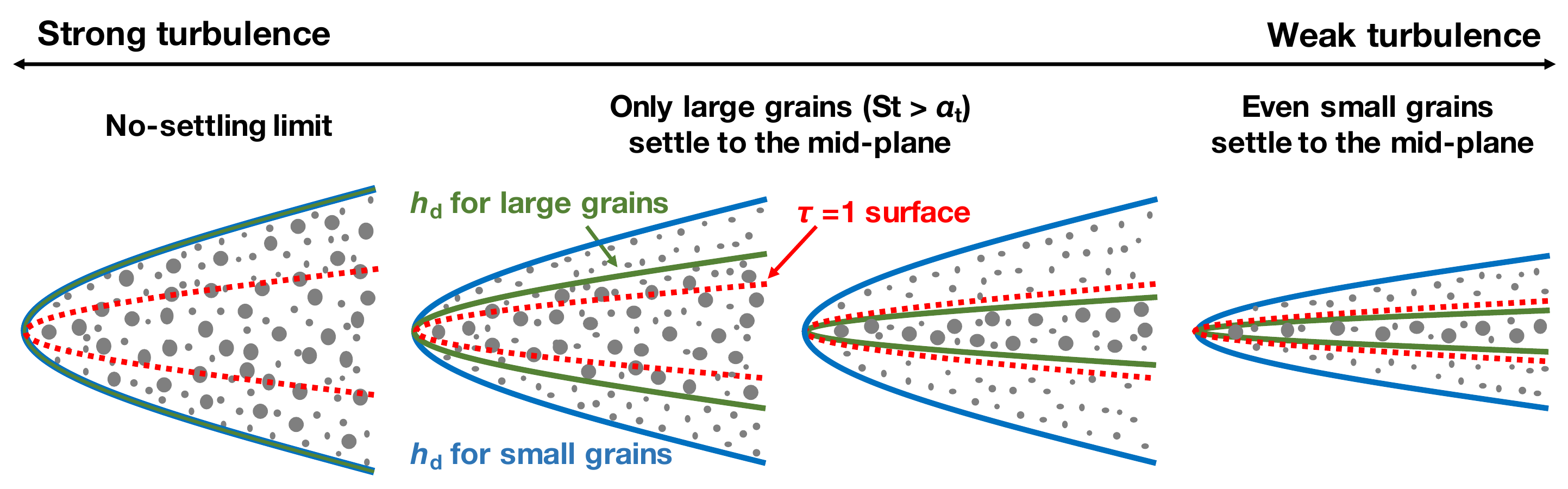}
\caption{
Schematic of the location of $\tau=1$ surface relative to dust scale heights.
In the strong turbulence regime, only the dust scale heights of large grains whose ${\rm St}$ is larger than $\alpha_{\rm t}$ decrease with decreasing the turbulence strength.
At some point, most of grains settle to the mid-plane since even small grains satisfy ${\rm St}<\alpha_{\rm t}$.
}
\label{fig:schematic}
\end{center}
\end{figure*}

Interestingly, in the models of mm-sized grains, the simulated polarization fraction at ALMA Band 6 and 7 reaches the maximum when $\alpha_{\rm t}\sim10^{-5}$ and cannot be higher even if the turbulence strength gets lower than $10^{-5}$.
This is because, in the weak turbulence regime ($\alpha_{\rm t}\lesssim10^{-5}$), almost all grains settle to the mid-plane in the same manner.
Figure \ref{fig:occupancy-all} shows a breakdown of extinction optical depth coming from each dust size at radial position of 20 au of Fixed models of $a_{\rm max}=1000~{\rm \mu m}$ and $a_{\rm max}=2900~{\rm \mu m}$.
We clearly see that the breakdown of the optical depth coming from each dust bin does not change with the turbulence strength for very weak turbulence regime.
If $\alpha_{t}\ll{\rm St}\ll1$, the dust scale height is 
\begin{eqnarray}
h_{\rm d}\sim\sqrt{\frac{\alpha_{\rm t}}{\rm St}}h_{\rm g}.
\label{eq:dustheight2}
\end{eqnarray}
In this regime, the scale heights of the different dust population change with $\alpha_{\rm t}$ in the same way.
In other words, large grains cannot settle to the mid-plane with leaving small grains in the upper layer.
Therefore, the location of the $\tau=1$ surface relative to the dust scale height of each dust population does not change with $\alpha_{\rm t}$ (Figure \ref{fig:schematic}).
The critical dust radius above which the dust grain settles to the mid-plane ($\alpha_{\rm t}\sim{\rm St}$) can be estimated from Equation \eqref{eq:st}:
\begin{eqnarray}
a_{\rm crit}\sim11.4
\left( \frac{\rho_{\rm int}}{1.675~{\rm g~cm^{-3}}} \right)^{-1}
\left( \frac{\Sigma_{\rm g}}{300~{\rm g~cm^{-2}}} \right)
\left( \frac{\alpha_{\rm t}}{10^{-5}} \right)~{\rm \mu m}.
\label{eq:acrit}
\end{eqnarray}
From Equation \eqref{eq:acrit}, small grains can be left in the upper layer only if $\alpha_{\rm t}\gtrsim10^{-5}$. 

From the polarization analysis, we can exclude the possibility that the maximum dust radius is $>3$ mm at the center part of the HL Tau disk. 
Even though the model with $a_{\rm max}=160~{\rm \mu m}$ predicts higher polarization fraction than the observations, we cannot exclude the possibility that the maximum dust radius is 160 ${\rm \mu m}$ because the other polarization machanisms might take place and reduce the polarization fraction due to self-scattering in the observing wavelength.
However, by considering both the SED and polarization, we conclude that the maximum dust size is $\lesssim 1~{\rm mm}$ and the turbulence strength parameter is $\lesssim 10^{-5}$.

\section{Discussion}\label{sec:diss}

\subsection{Is turbulence indeed very weak?}
Our results show that the turbulence strength needs to be very low ($\alpha_{\rm t}\lesssim10^{-5}$).
The efficient dust settling in the HL Tau disk is consistent with the indication from the geometrical thickness of the rings at the outer region \citep{Pinte2016} and the previous SED analysis \citep{Kwon2011,Kwon2015}.
Although we put a upper limit on the turbulence strength as $\alpha_{\rm t}\sim10^{-5}$, the lower limit would be potentially set from the infrared observations.
\citet{Kwon2011} has shown that the HL Tau disk needs small grains that are vertically well mixed with gas to explain its high mid- and far-infrared emission.
From Equation \eqref{eq:acrit}, dust grains smaller than 10 ${\rm \mu m}$ also settle to the mid-plane if $\alpha_{\rm t}<10^{-5}$, indicating that $\alpha_{\rm t}$ needs to be $\sim10^{-5}$ not well below $10^{-5}$.
The near-infrared scattered light observations would be helpful to constrain how much small grains are mixed up.
For example, the infrared observation toward the HD 163296 disk has shown that the outer region of the disk needs very weak turbulence corresponding to $\alpha_{\rm t}\sim10^{-5}$ to explain the non-detection of the scattered light \citep{Muro-Arena+18}.
However, the characterization from the scattered light is difficult for the HL Tau disk because it is still embedded in a massive envelope which prevent us from investigating the disk surface structure \citep{Beckwith+86,Beckwith+90}.

The very weak turbulence is consistent with the recent non-ideal MHD models where the Magneto-Rotational Instability (MRI) is suppressed at the disk mid-plane (e.g., \citealt{Gammie1996, Bai2017}).
However, even if MRI is suppressed, pure hydrodynamic instabilities would induce turbulence and lift up dust grains (e.g., \citealt{Flock2017, Flock2020}).

The gas surface density is also important parameter for the dust settling, although we fixed it as $1000~(r/{\rm au})^{-0.5}{\rm g~cm^{-2}}$.
As shown in Equation \eqref{eq:dustheight}, the settling behavior is determined by the ratio of ${\rm St}$ to $\alpha_{\rm t}$ and hence by $\alpha_{\rm t}\Sigma_{\rm g}$.
Therefore, if the gas surface density is 10 times lower than our model, 10 times higher $\alpha_{\rm t}$ is acceptable.
If the gas surface density is 10 time lower than our model, the dust-to-gas mass ratio is an order of 0.3 or higher.
In this case, the streaming instability might be operating and dust grains might be converted into planetesimals (e.g., \citealt{YG2005}).
If the dust-to-mass ratio is enough high, turbulence can no longer lift dust grains due to the drag force from dust to gas, which also helps the streaming instability takes place at the mid-plane \citep{Lin2019}.

\subsection{The impact of the other polarization mechanisms}
As mentioned above, this study focuses only on the polarization induced by self-scattering .
However, in observed disks, the other polarization mechanisms would operate and potentially cancel out the self-scattering polarization.
Our models with $a_{\rm max}=1~{\rm mm}$ and $\alpha_{\rm t}\lesssim10^{-5}$ predict polarization fraction comparable to the observations at ALMA Band 6 and 7, while it is significantly higher at ALMA Band 3 where the observed value is less than $0.1$\%.
This indicates that the the polarization fraction originating from the other mechanisms should be negligible at $\lambda\sim1~{\rm mm}$ and $\gtrsim0.3$\% (the polarization vector should be perpendicular to that from self-scattering) at ALMA Band 3.
The polarization induced by dust alignment is also sensitive to the dust size \citep{Guillet2020}.
\citet{Guillet2020} has shown that the alignment-induced polarization can be more effective at ALMA Band 3 than at ALMA 6 and 7 if $a_{\rm max}\sim 250~{\rm \mu m}$, while it is less sensitive to the observing wavelength if $a_{\rm max}\sim 1~{\rm mm}$.
This implies that if both self-scattering and alignment are taken into account, the true maximum dust size might be between 100 ${\rm \mu m}$ and 1 mm. 

The non-uniformity of the polarization at Band 6 indicates that the alignment-induced polarization takes place at this wavelength \citep{Stephens2017}, which seems to be in conflict with our mm-sized dust model.
However, in our model, we focus only on the optically thick inner region where the differential settling has a strong impact on the scattering-induced polarization.
The optical depth would decreases with the radial distance and hence the differential settling has a smaller impact at more outer region. 
This indicates that the polarization mechanism other than the self-scattering can dominate the polarization at the outer region, while self-scattering dominates at the inner region.

In addition to these uncertainties, the dust shape and internal structure also affect the polarization efficiency as well as the alignment efficiency of dust grains \citep{Tazaki2019,KBF2019,KB2020,Guillet2020}, which makes the interpretation of the disk polarization more complicated.
The comprehensive study including these effect would be necessary to understand the complex polarization behavior from the HL Tau disk.

\subsection{Effect of vertical temperature structure}
In our simulations, the temperature structure is assumed to be vertically isothermal.
However, the vertical temperature structure in protoplanetary disks is not necessarily isothermal in the vertical direction and might have an impact on the SED.

If the disk is passively heated by the central star, the temperature of the disk layer above the absorption surface for the stellar light would be higher than that of the layer below the surface.
The absorption surface of the disk is typically $\sim$ 4--5 scale height of the disk which is enough above the layer where the ALMA and VLA observed.
Therefore, the assumption of the vertically isothermal temperature structure would be valid if the HL Tau disk is passively heated.

If the disk is heated by gas accretion, the disk interior has higher temperature than the upper layer (e.g., \citealt{Chiang1997}).
The internal heating affect the SED since the longer observing wavelength traces the hot lower layer \citep{Sierra2020}.
The observed SED shows that the intensity at ALMA Band 4 is lower than the intensity extrapolated from the intensity at shorter wavelengths with the spectral slope of 2.
This indicates that the internal heating is not so effective otherwise the SED needs extremely strong intensity reduction at ALMA Band 4.

\subsection{Polarization at (sub-)cm wavelengths}
Our results showed that the observed polarization fraction can be explained with the maximum dust size of $1000~{\rm \mu m}$ but $160~{\rm \mu m}$-sized grains might be also acceptable if the other polarization mechanisms operate.
Our results also showed that the polarization fraction at $\lambda=8~{\rm mm}$ is 0.1\% for $a_{\rm max}=160~{\rm \mu m}$, while it is $\sim$ 4\% for $a_{\rm max}=1000~{\rm \mu m}$.
This is because the polarization efficiency has a peak at $\lambda\sim2\pi a_{\rm max}$ \citep{Kataoka2015} and the dust settling has a little impact on the polarization at $\lambda\sim {\rm cm}$ owing to the low optical depth.
Therefore, the polarimetric observation at (sub-)cm wavelengths using e.g., ngVLA, will be a good tool to solve the degeneracy and to know true dust sizes in protoplanetary disks.

\subsection{Radial profile of the dust size and turbulence strength}
In this paper, we focus only on the intensity at the center of the disk images, but it would be worth to make reference to the radial distribution.
\begin{figure}[ht]
\begin{center}
\includegraphics[scale=0.6]{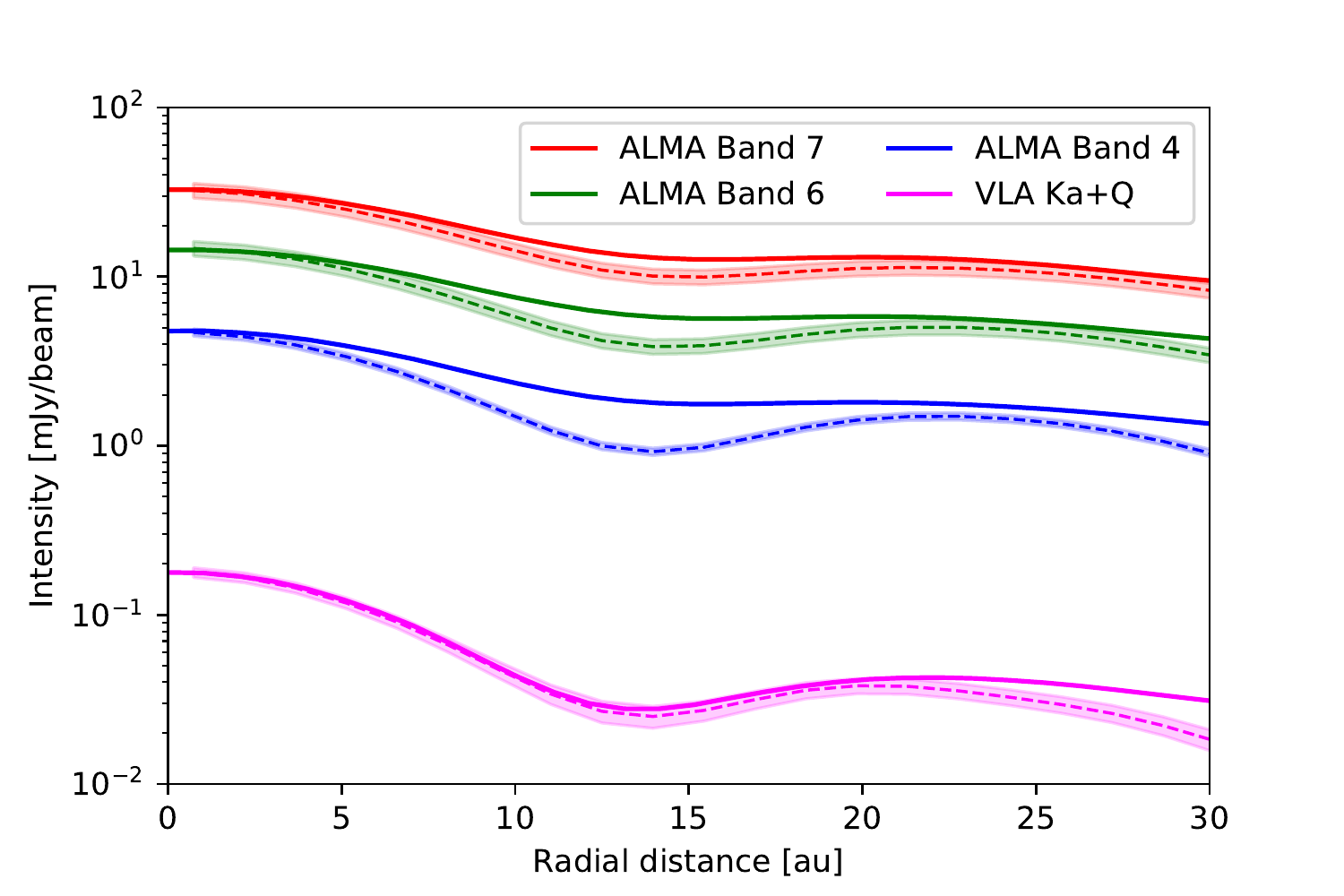}
\caption{Azimuthally averaged intensity profile of the HL Tau disk.
The solid line shows the intensity obtained from our turned model with $a_{\rm max}=1000~{\rm \mu m}$ and $\alpha_{\rm t}=10^{-5}$.
The dashed line shows the observed intensity profile.
The systematic error potentially caused in the calibration process is denoted with the translucent color.
For the VLA observation, the thermal noise is also taken into account in the error.
}
\label{fig:radial}
\end{center}
\end{figure}
Figure \ref{fig:radial} compares the radial intensity profile obtained in Tuned model ($a_{\rm max}=1000~{\rm \mu m}$ and $\alpha_{\rm t}=10^{-5}$) with observed one.
Our model predicts higher intensity at the gap region especially at ALMA Band 4.
This indicates that the dust size (and hence $\epsilon$) and/or turbulence strength in the gap would be different from our model.
To reduce the intensity ALMA Band 4, we need higher albedo than our model, suggesting that the turbulence strength is stronger than our model in the gap if the dust size is $a_{\rm max}=1000~{\rm \mu m}$.
Although detailed modeling of the radial intensity profile is out of our focus, it will be important to model the radial dependence for understanding how disk properties changes across the substructures.
The polarimetric observations with higher angular resolution with ALMA would be required to quantify the disk properties in more detail.

We also performed simulations without the gap to check the effect of the gap structure and found that the polarization fraction is slightly higher for the no-gap case.
This is because the no-gap model is optically thicker which makes the effective dust size smaller.
We confirmed that the slight difference due to the gap has a little impact on our results.

The HL Tau disk also shows the uniform polarization pattern even in the outer region ($\sim100~{\rm au}$), indicating that the dust size is $\sim 100~{\rm \mu m}$ for the entire region.
Although we demonstrated that the mm-sized grains can explain the observed polarization fraction at the center of the disk, it is not clear that the differential settling can explain the polarization observations in the entire region of the disk.
If the disk is optically thin, the differential settling no longer has an impact on the SED and polarization fraction.

\citet{Okuzumi2019} showed that the observed polarization pattern in the HL Tau disk can be explained by the fragmentation of non-sticky icy grains. 
However, the uniformity of the polarization degree needs a flat gas density profile because size of the fragments, which is determined through the Stokes number, needs to be uniform.
From our results, since the mm-sized grains can explain the observed polarization in the optically thick inner region, the gas density profile might not need to be flat.
The detailed modeling for the entire disk including the differential settling will give us a comprehensive understanding on the dust evolution in the radial direction.

\section{Summary}\label{sec:summary}
We performed radiative transfer simulations of disks with an analytical model of settling-mixing equilibrium of dust grains.
The simulated SED and polarization fraction were compared with the observations of the protoplanetary disk around HL Tau to constrain dust size and turbulence strength.

The SED of the center part of the HL Tau disk shows that the intensity slope between ALMA Band 6 and 7 is consistent with the spectral index of 2 within the error.
The observed intensity at ALMA Band 4 is below the value extrapolated from the intensity at ALMA Band 6 and 7 with the spectral slope of 2.
The models with the maximum dust size of $160~{\rm \mu m}$ can reproduce the observed SED only if $\alpha_{\rm t}\lesssim10^{-5}$, while the models with mm-sized grains can reproduce with a broad range of $\alpha_{\rm t}$.

The polarization analysis allowed us to constrain the turbulence strength more strongly.
If the maximum dust size is $160~{\rm \mu m}$, the polarization fraction is comparable or higher than the observed value at ALMA Band 6 and 7 for $\alpha_{\rm t}\gtrsim10^{-5}$.
If $\alpha_{\rm t}\sim10^{-6}$, the simulated polarization fraction at ALMA Band 7 is lower than the observed value.
The models with the maximum dust size of $1~{\rm mm}$ can explain the observed polarization fraction at both ALMA Band 6 and 7 if the turbulence strength parameter $\alpha_{\rm t}$ is $\lesssim10^{-5}$.
Although the observed polarization fraction at ALMA Band 3 is lower than that expected from the models, the other polarization mechanisms might reduce the scattering-induced polarization fraction.
If the maximum dust size is $3~{\rm mm}$, the simulated polarization fraction at ALMA wavelengths is significantly lower than the observed values.

To explain both the SED and polarization, the maximum dust size of $\lesssim 1~{\rm mm}$ and the turbulence strength parameter of $\lesssim 10^{-5}$ are required.
The efficient dust settling in the HL Tau disk is consistent with the previous studies (\citealt{Kwon2011, Kwon2015, Pinte2016}).
The degeneracy between $100~{\rm \mu m}$-sized grains and mm-sized grains can be solved by the polarimetric observations at (sub-)cm wavelengths using e.g., ngVLA.
These results showed that the differential settling has a key role in understanding the polarimetric observations on the optically thick inner region of disks.

\acknowledgments
We thank Cornelis P. Dullemond for an advice on the RADMC-3D simulations.
This work was supported by JSPS KAKENHI Grant Numbers JP18K13590, JP19J01929 and JP19H05088.
CC-G acknowledges support by UNAM DGAPA-PAPIIT grant IG101321 and CONACyT Ciencia de Frontera grant number 86372.
A.S. acknowledges support from ANID/CONICYT Programa de Astronomia Fondo ALMA-CONICYT 2018 31180052.
Numerical computations were in part carried out on Cray XC50 at Center for Computational Astrophysics, National Astronomical Observatory of Japan.
\software{RADMC-3D \citep{RADMC}}

\appendix

\bibliographystyle{aasjournal}
\bibliography{hltau}

\end{document}